# Black silicon avalanche photodiode with dopant-free multiplication region achieves >98% EQE

Oskari Leiviskä[1]

Olli Setälä[1]

Antti Haarahiltunen[2]

Juha Heinonen[2]

Ville Vähänissi[1]

Hele Savin[1]

[1]Department of Electronics and Nanoengineering, Aalto University, Tietotie 3, 02150 Espoo, Finland

[2]ElFys Inc., Tekniikantie 12, 02150 Espoo, Finland

Correspondence: Oskari Leiviskä

Aalto University, Department of Electronics and Nanoengineering

Tietotie 3, FI-02150 Espoo, Finland

E-mail: oskari.leiviska@aalto.fi

**Abstract:**

Conventional silicon avalanche photodiodes (Si APDs) rely on highly doped regions to enable impact ionization and achieve internal gain. However, inefficient charge collection in these regions, combined with front surface reflection, degrades the external quantum efficiency (EQE) of these devices. In this work, we mitigate both loss mechanisms by integrating black silicon (b-Si) surface nanostructuring with $Al_2O_3$ induced carrier collection in an inverted-mesa Si APD architecture. This design confines the electric field and defines the multiplication region without requiring ion implantation. The resulting devices display near-ideal responsivity (at M = 1) across the UV–visible spectrum, with EQE exceeding 100% at 200–400 nm (peak ~130%) and exhibiting 92–100% at 400–700 nm. Avalanche gains of M ≈ 4 at 80 V, M ≈ 10 at 100 V, and M ≈ 23 near breakdown at ~110 V are obtained. Dark current remains in the picoampere range up to ~80 V but rises to the nanoampere range near breakdown, narrowing the practical bias window. Capacitance-limited rise times of ~30–570 ns are measured for device diameters of 1–5 mm. The results demonstrate that the developed architecture enables APDs capable of detecting every single photon over wide range of wavelengths.

# Introduction

Silicon (Si) avalanche photodiodes (APDs) are of great interest to a wide range of high-sensitivity photodetection applications, including medical imaging[1], visible light communication (VLC)[2,3], light detection and ranging (LiDAR)[4], and defense[3]. However, their current external quantum efficiencies (EQE) remain sub-optimal because of conventional device architectures suffering from optical losses (i.e., surface reflection) and electrical losses (i.e., charge carrier recombination). Especially in the UV wavelengths of 200–400 nm, where Si absorption depth is only a few nanometres to ~100 nm[5], the EQE of commercial Si APDs is far from ideal, typically ranging from <10% to ~60%[6–8]. This is because it is difficult to achieve low reflection in the UV with conventional AR-coatings[9], and because surface ion-implantations introduce Auger recombination and non-depleted layers[10] that degrade the collection efficiency for charge carriers generated near surface.

Concurrently, major improvements for Si photodiodes (PDs) have been achieved by pairing surface nanostructuring (i.e., black silicon, b-Si) to suppress optical losses with a dopant-free charge collection layer induced by a charged dielectric film, typically atomic layer deposited (ALD) aluminium oxide ($Al_2O_3$), to minimize electrical losses[11,12]. These devices are commonly referred to as b-Si induced junction photodiodes or b-Si charge-induced electric field photodiodes in the literature[11–13]; here, for simplicity, we refer to them as b-Si I-PDs. The photosensitive area in these b-Si I-PDs is nanostructured into randomly distributed needles by inductively coupled reactive-ion etching (ICP-RIE), though non-plasma etching methods can be also used[14]. Such surface morphologies can achieve broad-spectrum reflectance well below 1% ranging from UV to near-infrared (NIR)[15,16]. The nanostructured surface of b-Si I-PDs is coated with an $Al_2O_3$ thin film. The fixed charge at the Si/$Al_2O_3$ interface[17] induces an electric field that is used to separate and collect the generated charge carriers[11], similarly to a conventional impurity-doped pn-junction. However, unlike in pn-junctions realized by implantation, the formation of dead layer at the very surface of the device that hinder the UV-performance can be avoided, allowing better carrier collection[11,12]. This together with the excellent surface passivation provided by the same $Al_2O_3$ thin film and the below 1% reflectance enabled by the b-Si results in close-to-ideal EQE values. Furthermore, these b-Si I-PDs have shown even above 100% EQE values in the UV region, where generated primary electrons have enough energy to trigger secondary carrier ionizations before recombining[11,12]. These excellent results from the earlier PD studies merit the investigation of adapting b-Si together with the charge induced electric field into Si APDs as well. So far, the induced electric field concept and b-Si have received little attention in APDs; prior studies have focused only on modestly improving NIR responsivity via femtosecond (fs) laser b-Si texturing[18,19].

In this paper, we explore the idea of converting the b-Si I-PD into an APD, to achieve internal avalanche gain together with close-to-ideal EQE. To realize this, we propose a new Si APD architecture utilising b-Si and induced electric field -based charge collection together with an inverted mesa design for local confinement of the multiplication region. First, we evaluate the feasibility of the proposed design by using TCAD simulations to estimate the electric field profiles, followed by a fabrication of a batch of b-Si charge-induced electric field avalanche photodiodes (I-APDs). We then assess their performance through EQE, gain and speed measurements, and compare the results with existing b-Si I-PDs and commercial Si APDs.

## *Design of b-Si I-APD Concept*

To convert the b-Si I-PD into an avalanche photodiode, it is necessary to achieve an electric field exceeding the critical value (typically ~2–5×$10^5$ V/cm in Si)[20] needed for sustaining avalanche gain (impact ionization). To evaluate if such a high field is possible to achieve in a conventional b-Si I-PD, we first study the electric field distribution in such a device under high bias voltage using Silvaco Atlas TCAD simulations. Fig. 1a shows the simulated p-i-n type I-PD structure, where the $p^+$ and $n^+$ contacts are formed on opposite sides of an $n^-$ type wafer using ion implantation, while the photosensitive b-Si surface region remains non-implanted. As described in the introduction, charge separation and collection within this region is ensured by forming an electric field through the deposition of an $Al_2O_3$ thin film with a high density of negative fixed charge.

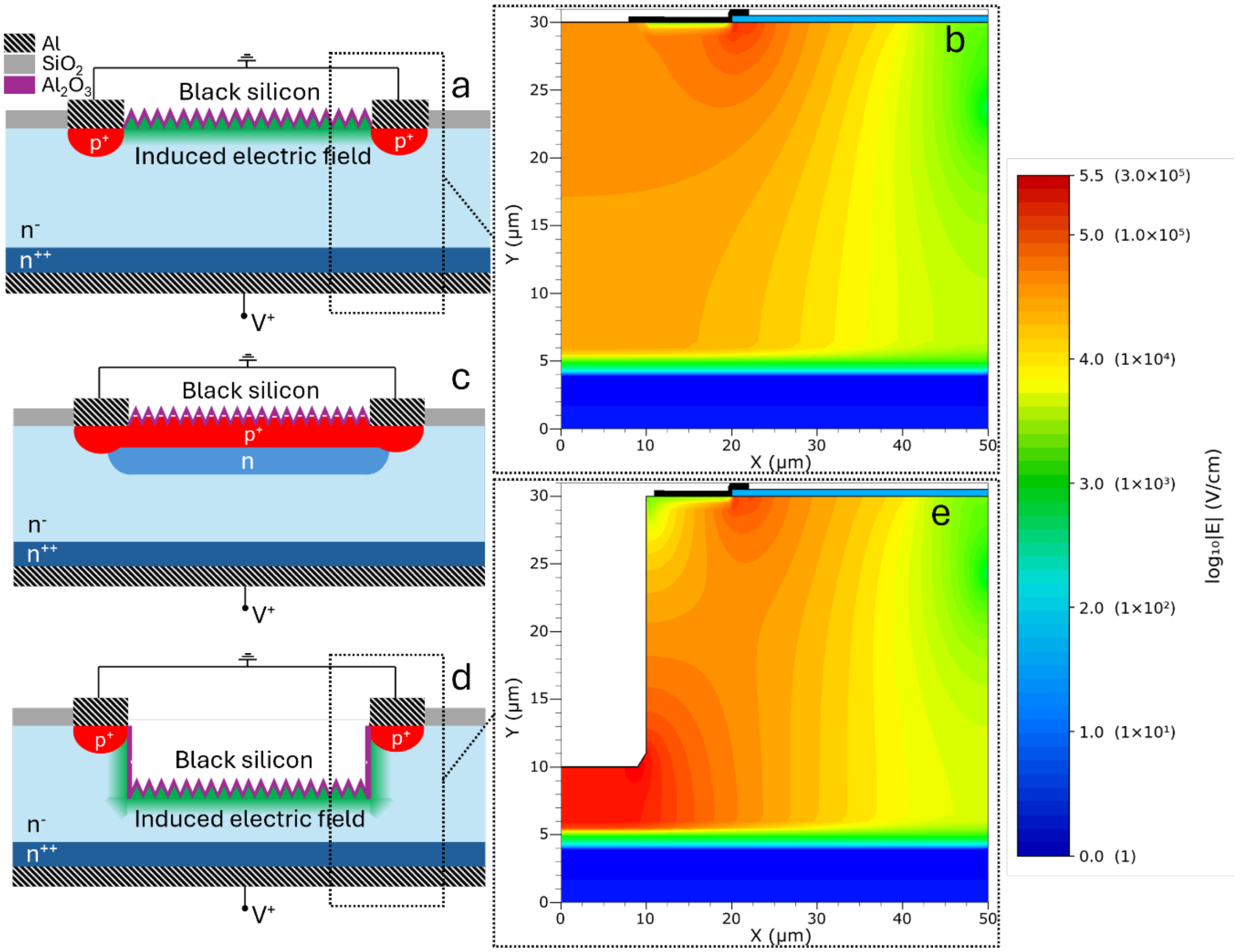

*Fig. 1.* ***b-Si APD concepts*** *Simplified schematic of (a) I-PD with b-Si and $Al_2O_3$ fixed charge induced carrier collection (shown as green) with ion implanted $p^+$ (shown as red) only below metal contacts and (b) simulated electric field magnitude ($log_{10}|E|$) near the contact edge and photosensitive region of reverse biased (100 V) of b-Si I-PD device shown in Fig. 1a, (c) traditional APD with b-Si fabricated on top of ion implanted $p^+$ active region, (d) locally thinned b-Si I-APD design with ion implanted $p^+$ only below metal contacts, and (e) simulated electric field magnitude of reverse biased (100 V) b-Si I-APD device shown in Fig. 1d.*

In the simulations, few simplifications were made to avoid an overly complex mesh; the b-Si surface morphology was omitted, and axial symmetry about the y-axis was exploited so that only half of the device was simulated. The effect of b-Si was incorporated solely through increased fixed charge density at $Si/Al_2O_3$-interface. Interface charge densities of $-2.5\times10^{12}$ and $-2.25\times10^{13}$ $cm^{-2}$ were used for planar and b-Si surfaces, respectively[21].

Even without simulations, it is evident that simply increasing the reverse voltage bias to increase the electric field in the structure of Fig. 1a to achieve controlled avalanche is problematic. With typical wafer thicknesses being in the range of a few hundred micrometres, reaching the critical field would require thousands of volts of reverse bias. Moreover, the increased bias voltage not only increases the electric field magnitude but also expands the depletion region horizontally into the substrate, increasing unwanted collection of dark current.

To keep operating voltages reasonable and mitigate horizontal field spreading, the $n^-$ region can instead be realised with moderately thin epitaxial layer (a few µm to tens of µm), which enables lower operating voltages and improves the field confinement. However, another issue would emerge by simply voltage biasing the existing I-PDs, even if a thinner epilayer would be used. The applied voltage would increase the electric field more below the doped $p^+$ ohmic contact than on the photosensitive area due to localised field enhancements (i.e., field crowding) arising from two abrupt field gradients caused by dopant-profile curvature at the $p^+$ edge[22,23] and by positive oxide charge at the $Si/SiO_2$-interface beneath the Al metal edge[23,24]. This field enhancement would lead to premature edge breakdown outside the intended photosensitive area as seen in the simulation (high field region {x, y=20, 30} in Fig. 1b), thereby posing challenges for proper APD functionality. In conclusion, thin epitaxial layer is not a proper solution either and different device design is needed.

Traditionally, the electric field is controlled in both lateral and vertical directions using doped wells and guard rings, which confine the peak electric field to the intended multiplication region and set its magnitude; this prevents premature breakdown and allows control of parameters like excess noise factor and gain[25]. In addition, ultra-shallow p+ surface junctions have been used to place the high field near the surface for improved UV response[26]. For our b-Si I-APD, an implantation-based approach could be done by forming an n-well with a $p^+$ surface region by ion implantations through the b-Si nanostructure as shown in Fig. 1c. However, adding doped regions to our device would negate the benefits of the original surface design inherited from the I-PD (i.e., no surface dead layers). Although shallow boron implantations through b-Si nanostructures can in some cases be done while avoiding excess recombination and creation of dead layers[27,28], more complex implantation processes would be needed for APDs. The strict requirements for the multiplication region, such as precise dopant profile control and minimal carrier recombination are extremely difficult to achieve by doping through the uneven b-Si surface. Therefore, the conventional multiplication region design is not suitable for our study either.

To achieve gain while preserving the dopant-free charge collection mechanism of the b-Si I-PDs, we propose a design in which the electric field magnitude is increased locally and the avalanche region is separated horizontally from the rest of the device by locally thinning the photosensitive region – effectively forming an inverted mesa in this area (see Fig. 1d). With this locally thinned design, the electric field increases more within thinned regions than non-thinned regions as bias is applied and the horizontal spread of the field outside the desired area is also minimised. Therefore, the electric field achieves the critical value to allow avalanche multiplication at the thinned active region (i.e., the region with black silicon and charge induced inversion layer) before impact ionization begins to occur in other non-active regions of the device. The simulated electric field of the proposed inverted-mesa design in Fig. 1e shows that mesa thinning can indeed confine a near-uniform critical field within the thinner epitaxial layer, while avoiding premature breakdown at ohmic contact edge due to local field crowding.

Based on the above simulations, we decided to build our b-Si I-APD design on epitaxial high resistivity n-type silicon (on a highly doped n-type carrier wafer) which is thinned down locally only from the active photosensitive region as shown in Fig. 2a. The b-Si nanostructure is formed on the bottom of this thinned inverted-mesa active region. The whole inverted-mesa structure is finally passivated by depositing an $Al_2O_3$ thin film covering sidewalls and the b-Si surface on the bottom. The high negative charge at the Si/$Al_2O_3$-interface, generates an electric field that separates generated charge carriers. It also attracts holes towards the surface, forming a $p^+$ inversion layer that serves as carrier-transfer channel along the sidewalls and bottom, thereby distributing the applied bias across the entire inverted mesa.

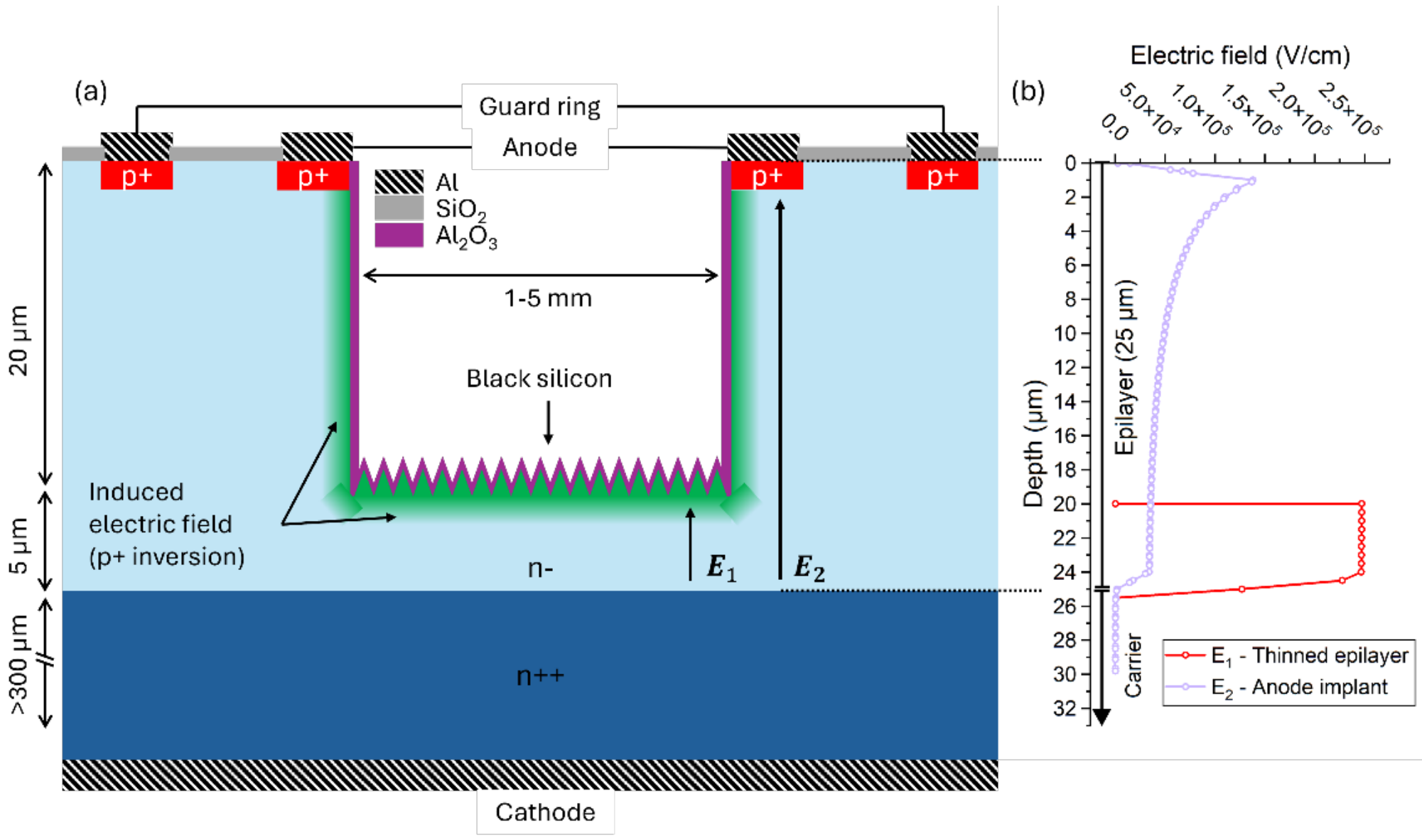

*Fig.2* ***Device cross-section and 1D-field cutline*** *(a) Cross-section of the proposed b-Si I-APD design. (b) Simulated 1D electric field magnitude inside the thinned 5µm epilayer with implantation free induced field collection* $|E_1|$ *and below the ion implanted* $p^+$ *anode* $|E_2|$ *at 120 V reverse bias, showing that the entire thinned epilayer operates as the multiplication region at high reverse bias.*

The above design allows us to tailor the electric field strength within the avalanche region at specific operating voltages by altering the etching depth of the inverted mesa, as the field magnitude is proportional to the layer thickness. Based on the simulations, thinning 25 µm $n^-$ type (5k Ω·cm) epilayer down to a 5 µm thickness was found to produce avalanche gain across the thinned region before ionization begins to dominate underneath the ohmic contacts (see Fig. 2b), and therefore this thickness was selected for the empirical work.

## Results & discussion

First, we quantified the external quantum efficiency of our b-Si I-APD at unity gain, providing a baseline for the comparison with I-PDs and other Si APDs. Fig. 3 shows the measured quantum efficiency at zero voltage bias, alongside reference data from b-Si I-PD and selected commercial Si APDs. The results show that our b-Si I-APD achieves near-ideal EQE performance over a wide range of wavelengths, like previously reported for b-Si I-PDs[11–13], which proves that the charge collection mechanism of the induced electric field functions well regardless of the steep walled inverted-mesa structure. Fig. 3 also shows the reflectance of our b-Si I-APD, which is extremely low over the whole 280–1000 nm wavelength spectrum. Furthermore, spatial uniformity of reflectance in the active region under 656 nm laser illumination was measured to be 0.47±0.25% (N=416).

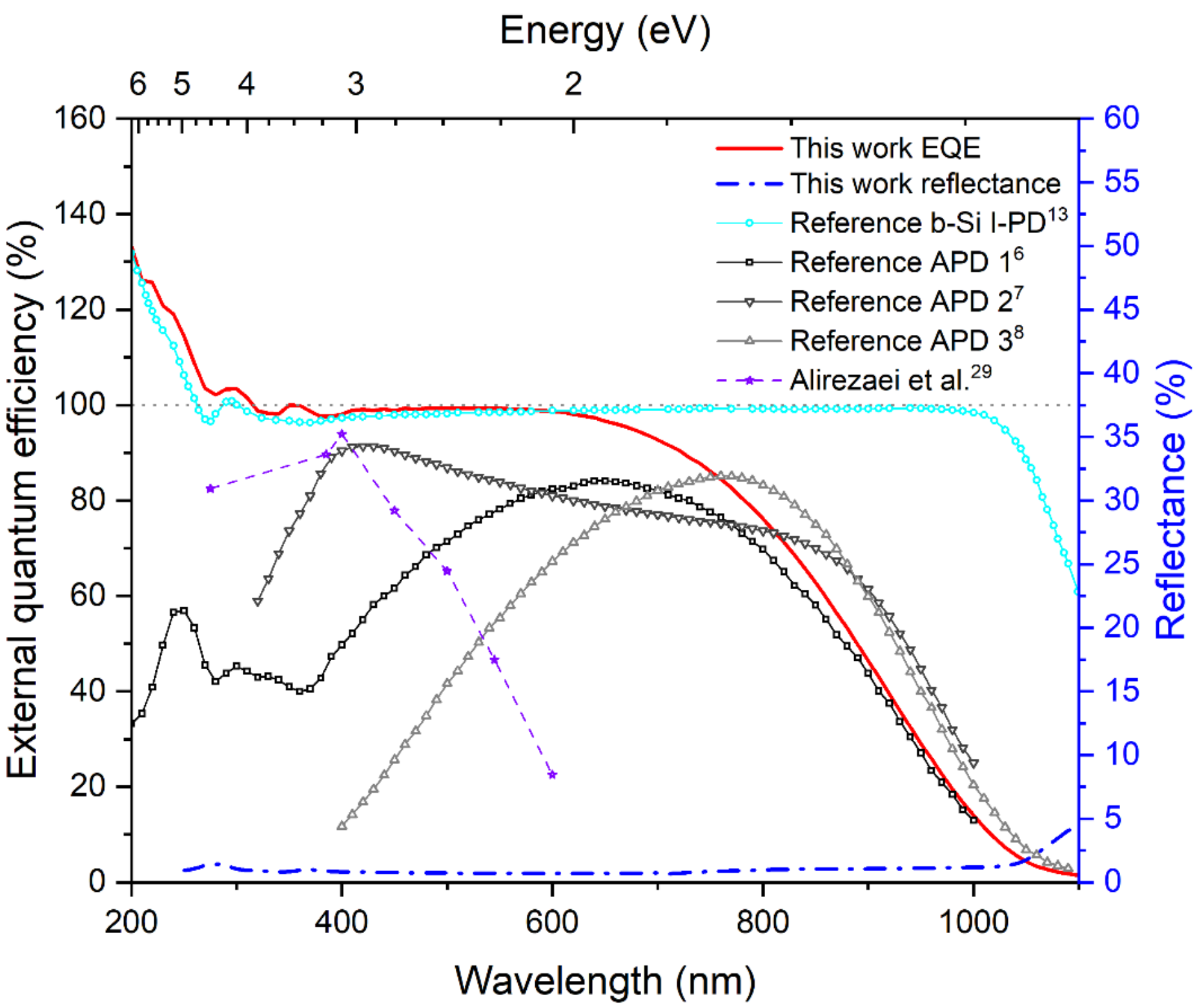

*Fig. 3.* ***Measured EQE and reflectance*** *Zero bias external quantum efficiency and total reflectance of our b-Si I-APD. For comparison, digitized EQE data from published b-Si I-PD*[13]*, three commercial Si APDs*[6–8]*, and a non-commercial Si APD optimized for UV*[29] *are also presented.*

At wavelengths between 200 and 400 nm, the observed EQE values exceed 100% and even reach up to 133.2%, which is consistent with the behaviour of b-Si I-PDs[11–13]. This result is expected, as the near-surface device region of our I-APD is like that of the b-Si I-PDs, except for the inverted-mesa topography.

Between 400 and 600 nm, the EQE of our I-APD settles to >98%, similarly to b-Si I-PDs, as the photon energy falls below the threshold for secondary ionization. This means that the device still captures nearly every photon and efficiently collects the photogenerated electrons, but it no longer produces additional charge carriers from the excess photon energy. Around 700 nm, the optical absorption depth becomes comparable to the thickness of the thinned ~5 μm epilayer[5], leading to decreasing EQE at longer wavelengths. This behaviour arises because the absorption shifts into the highly doped substrate, where the electric field rapidly vanishes, meaning carriers generated there rely on diffusion rather than drift. The high dopant concentration increases Auger recombination, shortening the minority carrier lifetime and thus the diffusion length. As a result, carrier collection becomes inefficient and the responsivity for NIR photons decreases. In contrast, the reference b-Si I-PD device with a thick (525 μm) lightly doped (>10 kΩ·cm) substrate allows longer diffusion lengths and efficient collection of charge carriers generated by deep-absorbing NIR light. Overall, across 200–600 nm, the EQE of our b-Si I-APD approaches ideal performance and matches state-of-the-art photodiodes.

For comparison, Fig. 3 shows the EQE of three commercial Si APDs: Reference APD 1 (Hamamatsu S12053[6]), Reference APD 2 (Hamamatsu S16453[7]), and Refence APD 3 (Excelitas C30902EH[8]) as well as one non-commercial Si APD reported by Alirezeai *et al.*[29]. The commercial devices are optimized for different wavelength regions (UV-Vis) and are representative of market-leading commercial performance, whereas the non-commercial device, to the best of our knowledge, represents state-of-the-art UV performance. It is immediately clear that the near-ideal EQE of our APD is much higher than reported for the commercial Si APDs at wavelengths <700 nm. Especially in the UV (200–400 nm), where the reflection tends to be more difficult to minimize with conventional anti-reflection coatings, our b-Si APD offers major improvement in the EQE. Even compared to the non-commercial state-of-the-art UV-enhanced Si APD reported by Alirezeai *et al.*[29] – which is based on a 400 nm thin Si body on silicon-on-insulator (SOI) technology – our APD achieves higher EQE in the UV, e.g., 102.8% at 275 nm and 98.0% at 400 nm, versus 82.8% and 93.9%, respectively. At longer wavelengths, our b-Si I-APD still outperforms the commercial Si APDs by retaining 99.4–92.8% EQE in the wavelength range of 400–700 nm. The reference APDs instead reach their peak EQE only around specific narrow wavelength regions, with peak values ranging from 82% to 91%. At longer wavelengths (>800 nm), Reference APD 3 outperform our device, owing to its thicker, NIR-optimized reach-through structure, achieving for example an EQE of 83.2% at 800 nm compared to 76.1% in our b-Si I-APD.

Because EQE at zero bias does not determine good APD performance by itself, the reverse biased operation of our device was characterized next to investigate its ability to produce avalanche gain, and to know how the dark current leakage behaves before device breakdown. Fig. 4 shows the measured dark current and photocurrent under blue LED light illumination together with the calculated photocurrent gain. When the reverse bias is swept from 0 to 80 V, the photocurrent remains constant, indicating unity gain, as in a conventional PD. Nevertheless, as we increase the voltage above 80 V, the photocurrent begins to increase. From the calculated gain, we can clearly see that the device starts to amplify the generated photocurrent above 80 V. This proves that our device with the induced electric field -based charge collection design can function as an APD. The increase of gain is more rapid after reverse bias reaches 100 V, as the gain is ~4 at 100 V and ~12 at 110 V. A peak gain of ~23 was measured at 112.5 V voltage bias after which further biasing resulted in a breakdown.

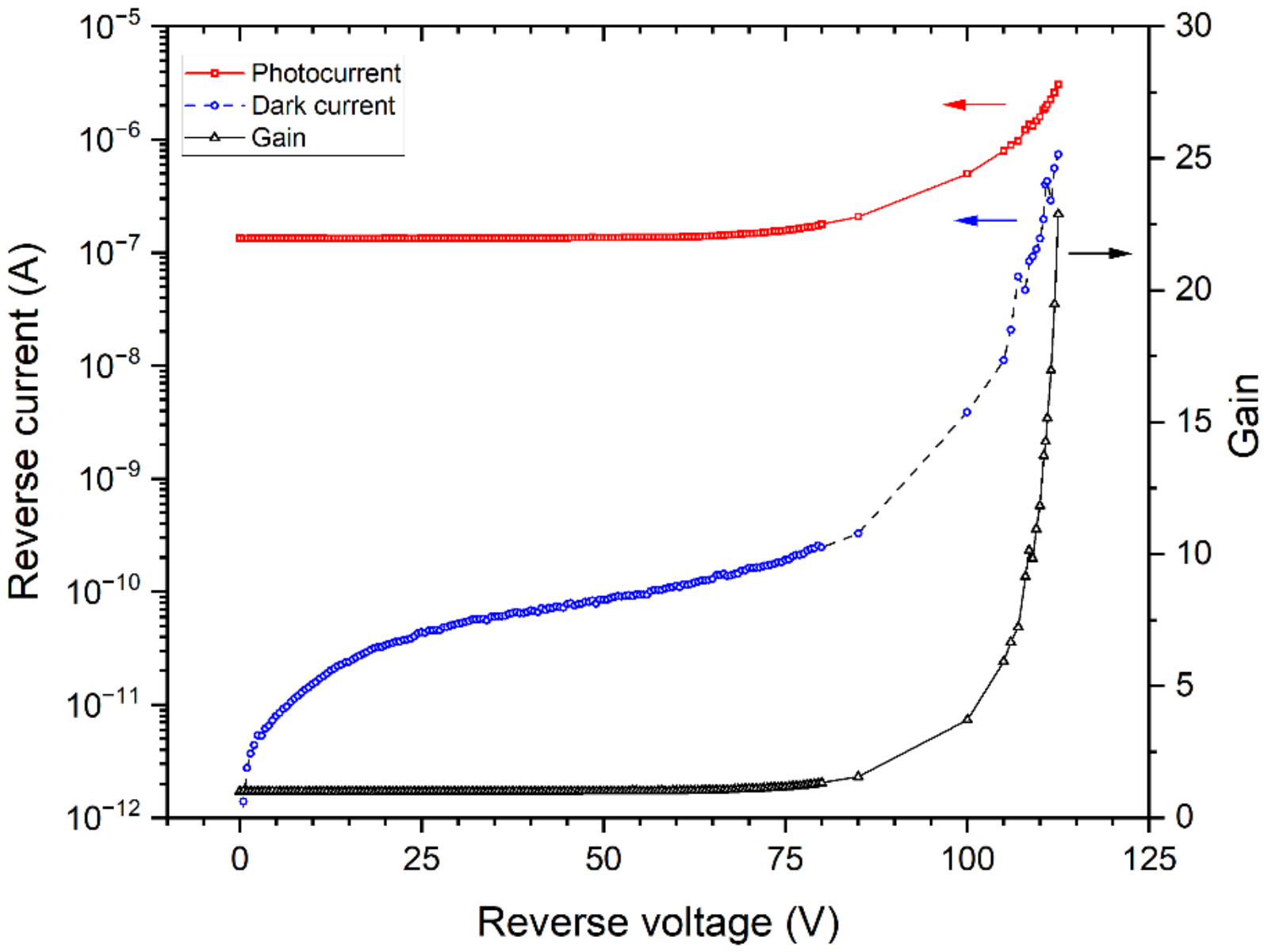

*Fig. 4.* ***Measured I-V characteristics*** *Measured photocurrent, dark current, and gain of our b-Si I-APD with a 3 mm active area diameter. During the illuminated measurement, a blue LED was used to provide a light injection with an arbitrary intensity. Gain was calculated as the ratio of the photocurrent at selected a voltage vs. at a unity gain (1 V).*

Dark current was measured to evaluate surface/bulk leakage introduced by local thinning and b-Si fabrication to verify the low-leakage operating range of our device. At 1 V reverse voltage the dark current is 2.8 pA (dark current density $J_d$=0.04 nA/cm$^2$) and it gradually increases to 245.8 pA ($J_d$=3.48 nA/cm$^2$) measured at 80 V reverse voltage. At higher reverse voltages, the dark current begins to increase very rapidly, and it is in the range of ~4–130 nA at bias voltages of 100–110 V, eventually reaching the level of the photocurrent. While avalanche gain naturally increases dark current alongside photocurrent, the dark current increase by a factor of >500 between 80 and 110 V is disproportionately large, suggesting additional contributions beyond multiplication alone. At high reverse bias, mechanisms such as trap-assisted and band-to-band tunnelling[30], microplasma breakdown due to nanostructure, and field-crowding at the edges of inverted mesa, may further enhance the leakage current. In comparison, the dark current of commercial Si APDs is typically in the order of a few nanoamperes at moderate gain levels[6–8].

The gain characterization above was performed under blue LED light illumination; however, gain in APD is expected to be strongly wavelength dependent due to the different generation depths of electron-hole pairs[31]. In silicon, electrons have higher impact ionization coefficient than holes, meaning that electron multiplication contributes more to the total avalanche gain[31,32]. The device should therefore produce peak gain when photoelectrons traverse the maximum distance within the multiplication region. Since carriers are generated at different depths, and since the vertical electric field in the avalanche region under reverse bias drives photoelectrons from the surface towards the bulk of the device (see Fig. 2a), short-wavelength photons (absorbed near the surface) generate electrons with longer path lengths in the multiplication region than long-wavelength photons absorbed deeper in the device. Consequently, short-wavelength illumination is expected to yield higher avalanche gain.

Fig. 5 shows the reverse bias dependent gain versus wavelength. As expected, the gain increases with bias. Under 280 nm illumination, the gain is ~4 at 80 V bias, ~10 at 100 V, and peaks at ~23 was using 107.6 V bias. Further increasing the bias to 107.9 V resulted in device instability, preventing further reliable measurements. It can be noted that the difference in breakdown voltage compared to the result shown in Fig. 4 (~108 V vs.

~113 V) indicates a small process variance (e.g., in epilayer thickness) between different devices from the same wafer.

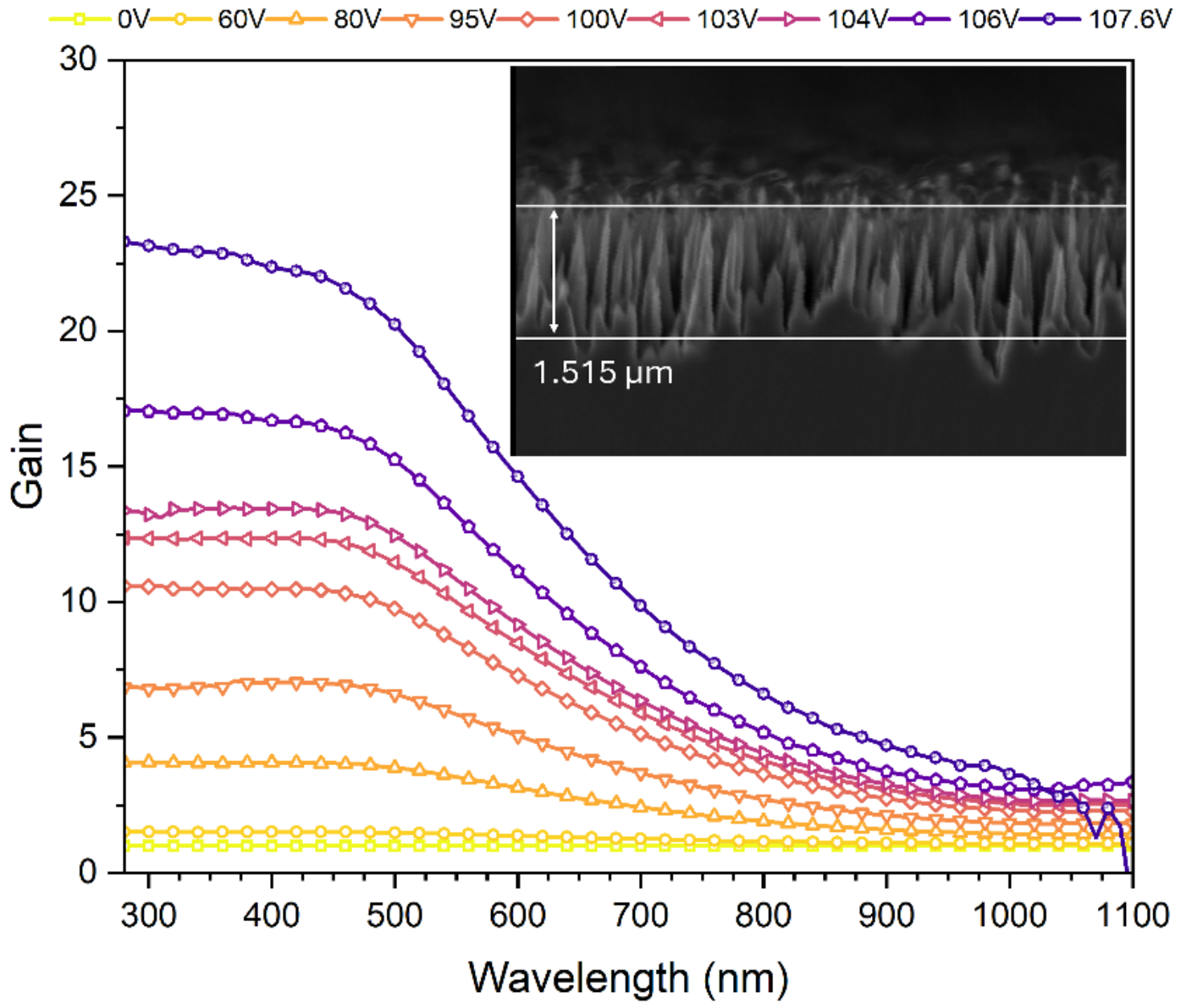

*Fig. 5.* ***Wavelength dependent gain*** *Calculated gain of our b-Si I-APD across a wavelength range of 280 to 1100 nm using different reverse bias voltages. Inset: Scanning electron microscope (SEM) image of b-Si nanostructures on the surface of the device demonstrating the thickness of the nanostructured layer.*

For photon wavelengths >450 nm, the gain begins to decrease with increasing wavelength at all measured bias voltages. This is well in line with the former assumption that gain should be lower as absorption depth increases, i.e., photoelectrons have decreased path length within the multiplication region. For wavelengths >700 nm, the absorption starts occurring mostly inside the $n^{++}$-bulk and the wavelength dependency of the gain weakens because the fraction of light absorbed within the first 5 µm multiplication layer remains rather constant. At 1000 nm wavelength, the EQE is only around 10% based on Fig. 3, as only a small fraction of light is absorbed within the first ~5 µm multiplication epilayer, decreasing the signal-to-noise ratio and measurement accuracy between 1000–1100 nm. Nevertheless, a gain of ~4 could still be observed at the highest voltage, demonstrating that avalanche multiplication remains detectable even under these conditions.

Assuming similar uniform electric field across the thinned multiplication layer as in simulations (see Fig. 1e and Fig. 2b), the gain should keep growing towards the shorter wavelengths also at 280–450 nm, as the path length of electrons in the high field region increases. However, the gain reaches a plateau for wavelengths below 450 nm for all voltages except the highest one (107.6V), which exhibits slightly increasing gain <450 nm. This indicates that the avalanche region does not extend to the very surface of the nanostructure, making the gain constant for wavelengths absorbed before it. The absorption depth for 450 nm photons is around 900 nm, which is in the same scale with the mean height of our b-Si needles (see inset in Fig. 5). This suggests that below 450 nm wavelengths, generation mostly occurs inside b-Si needles and the magnitude of the electric field inside the needles remains below the critical field. In other words, the multiplication region does not extend to the nanostructured surface, which prevents impact ionization inside the b-Si needles. In this case, photoelectrons generated inside the b-Si structure simply drift to the remaining epilayer without multiplication after which they begin to experience the gain. Thus, gain remains constant until the absorption begins to shift outside the b-Si needles at longer wavelengths. This idea was validated via preliminary simulations, which showed that the electric field in the bulk of b-Si needles remains almost unaffected by increased reverse bias voltage, but more systematic studies are still needed to prove this.

Finally, we measured the capacitance and rise time to ascertain that our device structure does not cause any issues on these metrics. For fully depleted devices (−10 V bias) with 1 and 5 mm diameters, the capacitances are 23 and 423 pF, respectively. By modelling the device as a parallel plate capacitor formed by the depleted thinned epilayer, these capacitance values correspond device thicknesses of 3.8 and 4.9 µm, respectively, with the difference further indicating slight thickness variation throughout the fabrication process. This capacitance level was expected as it follows from the active layer thinning and therefore the capacitance of our inverted mesa structure is also fundamentally higher than in some other designs such as reach-through APDs[32], which can utilize much thicker active regions with thicknesses up to few hundred µm. The impact of the high capacitance can also be seen in the rise times of the b-Si I-APD, which exhibit capacitance-limited (i.e., limited by the RC-time constant) values of 30 and 570 ns for 1 and 5 mm devices, respectively. Fundamentally, the design is not an optimal solution for ultrafast sensing, but the speed could be improved by decreasing the capacitance by downscaling the device diameter or by using a thicker epilayer, which would lead to increased operating voltages.

Overall, the architecture presented in this paper provides a viable transformation of the proven b-Si I-PD into a b-Si I-APD. Compared to the representative Si APDs, our device captures photons roughly 5–60 percentage points more efficiently across 400–700 nm spectrum, excluding the even higher differences in UV. Although the gain is only moderate compared to commercial Si APDs, the exceptionally high EQE increases the primary photocurrent and improves the signal-to-noise ratio without introducing additional noise. This is especially beneficial in shot noise –limited regimes where the improved EQE directly translates to better sensitivity whereas gain amplifies both signal and noise by a same factor. Furthermore, our lower avalanche breakdown voltage (~110 V) compared to the commercial references (~160–400 V)[6–8] allows lower operating biases. One drawback of our device is the dark current at high operating biases: while it remains relatively low up to 80 V (~3–246 pA), it rises quickly to ~4–130 nA in the 100–110 V range, which can exceed the commercial APDs[6–8]. Further dark current reduction is therefore needed to lower noise and to expand the stable bias range of our device.

## Conclusions

This work introduces a fundamentally new silicon avalanche photodiode architecture that overcomes two key limitations of state-of-the-art devices: 1.) carrier loss associated with highly doped regions and 2.) absorption loss resulting from high surface reflectance. Firstly, instead of relying on conventional ion implanted junctions, we demonstrate a dopant-free avalanche multiplication region formed by utilizing oxide charge and a mesa geometry to create a confined high electric field region. Secondly, we integrate black silicon nanostructuring onto the device, achieving below 1% broadband reflectance over 200–1100 nm. Our architecture enables high external quantum efficiency, achieving near ideal (>98%) EQE at 200–600 nm a peak of 130% at the UV end. Such a high broadband responsivity enables accurate detection across multispectral illumination with a single detector. When reverse biased to ~110 V, our device achieves a gain of 23, with the highest gain values recorded at the UV wavelengths. Dark current remains in the picoampere range up to ~80 V but rises to the nanoampere level near breakdown, which may narrow the bias window for the most sensitive operation. The APD device architecture developed here could be utilized for a wide variety of applications including UV–visible spectroscopy, environmental and space sensing, and next generation SPAD based imaging.

# Methods

## *Device Fabrication*

The processing of our Si APDs began with 150 mm diameter epitaxial silicon wafers (see Fig. 2a for the device cross-section). These wafers had a 25 µm high-resistivity (arsenic doped, $n^-$ type, 720–1680 kΩ·cm) epitaxial layer grown on a low-resistivity (antimony doped, $n^{++}$ type, 5–20 mΩ·cm) substrate. First, thick oxide layers were thermally grown on both surfaces to provide surface passivation on the non-active areas and to be used later as an implantation mask. After oxide pattering, boron and phosphorous were implanted on the front and rear surfaces, respectively, to ensure later formation of ohmic contacts between Si and aluminium (Al).

Next, a 20 nm $Al_2O_3$ hard mask was prepared using thermal ALD using trimethylaluminium (TMA) and $H_2O$ as precursors at 200 °C in a thermal-ALD reactor (Beneq, TFS-500). The $Al_2O_3$ hard mask was patterned with 1–5 mm diameter circular openings for the active region openings. The inverted-mesa of each device was formed by thinning the wafer locally on the patterned areas using cryogenic deep reactive-ion etching (DRIE), resulting in a ~5 µm thick remaining epilayer in these regions. After thinning, the black silicon nanostructure was etched with DRIE on the bottom of each inverted-mesa. These consecutive etching steps – thinning and b-Si formation – were done with inductively coupled reactive ion etching (ICP-RIE) system (Oxford Instruments,

PlasmaPro 100 Estrelas) using $SF_6$ and $O_2$ chemistry at −125 °C. The thickness of the remaining active epilayer was later confirmed by depletion capacitance characterization of the final device.

The surface passivation of the nanostructured active area and the formation of the charge collection channel were achieved by first removing the previous $Al_2O_3$ hard mask and re-depositing a new 30 nm thick ALD $Al_2O_3$ passivation layer on the photosensitive region using the same ALD process as in the hard mask preparation. After ALD, ohmic contacts were made by sputtering and patterning 300 nm of Al on both sides of the device. The annular anode and guard ring were 100 µm wide each. Finally, the $Al_2O_3$ surface passivation was activated in a 30-minute forming gas annealing at 425 °C temperature.

### *Device Characterization*

The EQE of our b-Si I-APDs was calculated from a measured spectral response $R_\lambda$ at zero bias (unity gain) as:

$$EQE(\lambda) = R_\lambda \frac{hc}{e\lambda} = \frac{I_{ph}}{P_{in}} \frac{hc}{e\lambda}$$

where, $I_{ph}$ is the measured photocurrent, $P_{in}$ is the incident optical power, $e$ is the elementary charge, $h$ is the Planck constant, $c$ is the speed of light, and $\lambda$ is the wavelength of the incident monochromatic light. These measurements were conducted by placing the detector in a dark box and illuminating them using monochromator (Bentham, TMC300) filtered white light from dual-lamp source (Bentham, ILD-D2-QH). The measurement was conducted with wavelengths between 200 and 1100 nm with 10 nm intervals. The EQE was initially calibrated against traceable photodetector (Newport, 818-UV) and output photocurrent from the sample was measured using Keithley 237 source meter unit. The sample temperature was controlled at constant value of 27 °C by integrated heating resistor on printed circuit board. With the former setup, the signal-to-noise ratio became a limiting factor in voltage biased responsivity at UV region measurements as dark current was higher. Therefore, the light source was replaced with xenon light source (Bentham, TLS120Xe), which provided better signal intensity in the UV. With this setup, the wavelengths were limited between 280 and 1100 nm in the wavelength dependent gain measurements.

The total reflectance of the b-Si texture in the final device was characterized using two complementary methods. First, reflectance mapping was performed to assess the uniformity of the surface. The mapping was conducted with a 656 nm laser, scanned across the detector surface with a 0.2 mm raster size (Semilab, PV-2000). Subsequently, spectral reflectance was measured over a wavelength range of 250–1100 nm using a monochromated dual light source coupled to an integrating sphere (Bentham, PVE300).

Current–voltage (I–V) characteristics of our APD were obtained at a probe station by measuring the current during varied reverse biasing between 0–113 V. These measurements were conducted inside a dark box and using a semiconductor parameter analyser (Hewlett-Packard, 4155A). A blue LED was used for crude light injection for a photocurrent measurement. To determine the gain during the I–V and spectral responsivity measurements, the corresponding currents were measured under both dark and illuminated conditions, from which the gain follows:

$$M = \frac{I_{ph}}{I_{ph,0}} = \frac{I_{ill} - I_{dark}}{I_{ill,0} - I_{dark,0}}$$

where $I_{ph} = I_{ill} - I_{dark}$ is the photocurrent at the measurement bias, and $I_{ph,0}$ is the primary (no-gain) photocurrent. $I_{ill}$ and $I_{dark}$ are the illuminated and dark currents at a specified bias voltage, while $I_{ill,0}$ and $I_{dark,0}$ are the corresponding illuminated and dark currents at unity gain (1 V), where avalanche multiplication is absent.

The capacitance–voltage (C–V) measurements were done with a low frequency impedance analyser (Hewlett-Packard, 4192A) using a 100 kHz AC signal. The reverse bias was increased until capacitance saturated and the active region thickness (i.e., for the locally thinned region) was derived from this depletion capacitance using parallel plate model:

$$d_1 = \frac{\epsilon_0 \epsilon_r \pi R_1^2}{C_{total} - \frac{\epsilon_0 \epsilon_r \pi (R_2^2 - R_1^2)}{d_2}}$$

where $C_{total}$ is the measured device capacitance, $R_1$ is the radius of thinned active region, $R_2$ is the outer radius of the annular anode contact, $d_2$ is the thickness of the pristine epilayer under anode, and $\epsilon_r$ is the relative permittivity of silicon. The horizontal spread of depletion outside of anode was neglected as suggested by the TCAD simulations.

The speed of our b-Si APD was characterized by measuring the rise time from its transient response to 1064 and 405 nm laser diodes using an oscilloscope (PicoScope, 3206D-PP961 200 MHz). The measurement setup was tested and monitored using fast reference photodiode (Vishay, BPV10) with 2.5 ns rise time (850 nm, 80 V). A detailed description of the setup is provided elsewhere[33].

## Author information

### *Corresponding author*

**Oskari Leiviskä**, Department of Electronics and Nanoengineering, Aalto University, Tietotie 3, FI-02150 Espoo, Finland; https://orcid.org/0009-0009-8897-8348; oskari.leiviska@aalto.fi

### *Author contributions*

A.H. conceived the initial concept. O.L. performed the TCAD simulations under the guidance of J.H. O.L. and O.S. planned and conducted device fabrication under the supervision of V.V. and H.S. O.L. and O.S conducted the I–V, C–V, and reflectance measurements. A.H. performed the zero-bias EQE and bias-dependent gain measurements. All authors contributed to the data analysis and interpretation. O.L. prepared the initial manuscript draft of the paper, which was reviewed and revised by all authors.

### *Acknowledgments*

The authors acknowledge Aalto University Micronova Nanofabrication Centre for providing facilities. We acknowledge the financial support of the Finnish Ministry of Education and Culture through the PREIN/I-DEEP doctoral pilot (VN/3137/2024-OKM-4). We also acknowledge the Flagship for Photonics Research and Innovation (PREIN) decision number Research Council of Finland/2024/368652. The work was funded through European Space Agency (ESA) under contract No. 4000144190/24/NL/AR/ahh (Avalanche Photo Diode based on Black Silicon (APD-bSi)).

### *Competing interests*

The authors declare the following financial interests/personal relationships which may be considered as potential competing interests: Juha Heinonen and Antti Haarahiltunen are shareholders and employees of ElFys Oy commercially providing black silicon photodiodes. The other authors declare no competing interests.